\documentclass[twocolumn,aps,pra,superscriptaddress,amsfonts]{revtex4-1}
\usepackage[colorlinks,linkcolor=blue,urlcolor=red,citecolor=red]{hyperref}
\usepackage{graphicx,amsmath,amsfonts,amssymb,capt-of}
\usepackage[tight]{subfigure}
\usepackage{yquant}
\usepackage{tikz}
\usetikzlibrary{quantikz}
\usepackage{siunitx}
\usepackage{bigints}
\usepackage{tabularx}
\usepackage{mathtools}
\usepackage{float}
\usepackage{multirow}
\usepackage{newfloat}
\usepackage{array}
\usepackage{cleveref}
\usepackage{tikz}
\usetikzlibrary{decorations.pathreplacing}
\usetikzlibrary{colorbrewer}
\usetikzlibrary{arrows}
\usetikzlibrary{arrows.meta}
\usetikzlibrary{calc}
\usetikzlibrary{positioning}
\usepackage{makecell}
\usepackage{comment}
\usepackage{physics}
\usepackage[utf8]{inputenc}
\usepackage[super]{nth}
\usepackage{amsmath}
\usepackage{geometry}
\usepackage{xcolor}
\definecolor{color1bg}{HTML}{8dd3c7} 
\definecolor{color2bg}{HTML}{ffffb3} 
\definecolor{color3bg}{HTML}{bebada} 
\definecolor{color4bg}{HTML}{fb8072} 
\definecolor{color5bg}{HTML}{80b1d3} 
\usepackage{bbm}
\newcolumntype{P}[1]{>{\centering\arraybackslash}p{#1}}

\usepackage{enumitem}
\usepackage{chemfig}
\usepackage[normalem]{ulem}
\usepackage{placeins}
\usepackage{changepage}
\newcolumntype{x}[1]{>{\centering\let\newline\\\arraybackslash\hspace{0pt}}p{#1}}
\DeclareFloatingEnvironment[
    fileext=loa,
    listname=List of Algorithms,
    name=ALGORITHM,
    placement=tbhp,
]{algorithm}

\makeatletter
\newcommand{\vast}{\bBigg@{4}}
\newcommand{\Vast}{\bBigg@{5}}
\newcommand{\VAst}{\bBigg@{6}}
\makeatother

\newenvironment{packed_item}{
\begin{itemize}
  \setlength{\topsep}{1pt}
  \setlength{\partopsep}{1pt}
  \setlength{\itemsep}{1pt}
  \setlength{\parskip}{0pt}
  \setlength{\parsep}{0pt}
}{\end{itemize}}

\usepackage{colortbl}
\usepackage{xr}

\makeatletter
\newcommand*{\addFileDependency}[1]{
  \typeout{(#1)}
  \@addtofilelist{#1}
  \IfFileExists{#1}{}{\typeout{No file #1.}}
}
\makeatother

\newcommand*{\myexternaldocument}[1]{%
    \externaldocument{#1}%
    \addFileDependency{#1.tex}%
    \addFileDependency{#1.aux}%
}

\myexternaldocument{SI}

\begin{document}
\title{Quantum Computing with dartboards}
\author{Ishaan Ganti}
\affiliation{Mission San Jose Highschool, Fremont, CA}
\author{Srinivasan S. Iyengar}
\email{Email: iyengar@indiana.edu}
\affiliation{Department of Chemistry, and the Indiana University Quantum Science and Engineering Center (IU-QSEC),
Indiana University, Bloomington, IN-47405}
\date{\today}

\begin{abstract}
We present a physically appealing and elegant picture for quantum computing using rules constructed for a game of darts. A dartboard is used to represent the state space in quantum mechanics and the act of throwing the dart is shown to have close similarities to the concept of measurement, or collapse of the wavefunction in quantum mechanics. The analogy is constructed in arbitrary dimensional spaces, that is using arbitrary dimensional dartboards, and for for such arbitrary spaces this also provides us a ``visual'' description of uncertainty. Finally, connections of qubits and quantum computing algorithms is also made opening the possibility to construct analogies between quantum algorithms and coupled dart-throw competitions. 
\end{abstract}

\maketitle

\section{Introductions}
The promise of solving exponentially complex problems efficiently using quantum technologies\cite{Preskill2018-NISQ,Nielsen-Chuang,Nai-Hui-Hybrid-QC} and the development of the associated software is a rapidly evolving research frontier. While we are in the early stages of this emerging quantum revolution, there is already a diverse set of problems that can benefit from such developments. However, true progress can only be achieved by rigorous study facilitated by the development of a competitive quantum workforce. As a result, research and workforce development in the strongly inter-disciplinary area of quantum information sciences has been noted by the US National Science Foundation (NSF) as one of the ``Big Ideas'' and recognized through the introduction of the National Quantum Initiative (NQI) from the White House.
Furthermore, the 2019 National Academies Report entitled \emph{Quantum Computing: Progress and Prospects}~\cite{NAS2019} observes that ``\emph{[a]dvances in QC theory and devices will require contributions from many fields beyond physics, including mathematics, computer science, materials science, chemistry, and multiple areas of engineering.}'' 
By contrast, regarding the present state of the Quantum Information Science (QIS) workforce, the September 2018 National Science and Technology Council report~\cite{NSTC2018} notes that ``\emph{America’s current educational system typically focuses on discrete disciplinary tracks, rarely emphasizing cross-disciplinary study that equips graduates for complex modern questions and challenges, prominently including QIS.}'' Emphasizing this point, Jeremy Hilton,
Quantum Engineering Lead at Google AI, 
wrote in Forbes~\cite{Hilton2019} that ``\emph{one issue has everyone united: There’s a shortage of quantum computing talent. This shortage has a significant impact on the future of the industry. A trained, well-rounded quantum workforce is key to realizing the full practical value of quantum computing. And yet, pundits describe time and again the difficulty in recruiting talent. There isn’t a direct pipeline from universities, and there’s fierce competition for the limited workforce that is available\ldots To continue expanding the quantum ecosystem, we need to grow the number of quantum-literate experts — now\ldots }''

Given these critical challenges it is vital that we develop new paradigms for quantum education that is accessible at multiple levels of pedagogy. Strongly influenced by other similar initiatives\cite{Rudolph-Q-is-for-Quantum,Rudolph-teaching-QC}, here we present an approach where we utilize a game of darts to discover the complexities of quantum mechanics and eventually quantum computing. The first author in this article is a high-school senior who has deeply benefited from this interdisciplinary initiative. In fact, the goal of this effort can be labeled as ``Quantum computing without physics." We make the case that it may be possible to introduce quantum theory without much background in physics at all. 

This paper is organized as follows. We will begin by discussing the Schrodinger's cat problem along with a two-dimensional game of darts with rules constructed to reproduce the cat-state problem. Then, we will generalize this game of darts to multiple dimensions and eventually to a Fourier space, which yields one visualization of uncertainty. Finally, we  connect the game of darts to measurements performed on a set of qubits states, and thus quantum computing, and analyze a simple quantum circuit by constructing an analogy to a game of darts.

\section{From cats to dartboards: setting the stage}
\label{cats}
Schr\"odinger’s cat is a central thought experiment detailing the following scenario: a cat is trapped in a box with a lethal device for a certain duration of time. During this time, the device has some probability of activation. Then, by the end of a fixed duration of time, the box is
opened and the state of the cat (alive or dead) is observed.

Schr\"odinger’s cat is as famous as it is because according to the principles of quantum mechanics, at the moment just
before the box is opened, the cat is in  a ``super-position'' state, that is, both dead and alive. And somehow, when the box is opened, the cat ``collapses'' to one of the two, dead or alive states. Of
course, this sounds extremely odd. Even Schr\"odinger thought so. Yet, if we consider this scenario as an analogy to the behavior of 
subatomic particles, this thought experiment accurately describes and presents the basis for much of contemporary non-relativistic physics, and also provides us with an algorithm to computationally inspect almost all phenomena in chemistry, biochemistry and material science.

To rationalize this idea and to set the stage for more general prescriptions, we may begin with a two-dimensional space where one axis is labeled as ``alive'', and the other is labeled as ``dead''. Any point in
this two-dimensional space represents the unobserved state of the cat prior to the box being opened. Thus we may say, 
\begin{align}
    cat = \alpha \cdot (alive) + \beta \cdot (dead)
\end{align}
or  formally,
\begin{align}
    \ket{cat} = \alpha \ket{alive} + \beta \ket{dead}
    \label{catstate}
\end{align}
This can be equivalently represented in a two-dimensional vector space as 
\begin{align}
\begin{pmatrix}
  \alpha \\
  \beta 
\end{pmatrix}
\label{cat-vector}
\end{align}
 assuming the first axis is the ``alive'' axis and the second is the ``dead'' axis.
 The only additional condition on choosing the specific point on the two-dimensional space is that the sum of the squares of the ``alive'' and
``dead'' components, that is, the length of the vector in Eq. (\ref{catstate}) 
must equal 1. In the Coppenhagen interpretation of quantum mechanics, this implies the probability that the cat is alive or dead is equal to 1. 
In other words, the probabilities of the cat being found ``alive'' or ``dead'' 
when the box is opened
correspond to $\alpha^2$ and $\beta^2$, respectively, and we have 
\begin{align}
\alpha^2+\beta^2=1. 
\label{norm}
\end{align}

\subsection{A Dartboard for Our Cat}
\label{darts}
Imagine a game of darts for the above scenario. We will define the rules of the game as follows:  
\begin{packed_item}
    \item[Rule 1] The action of throwing the dart is designed to correspond to the action of opening the box in Schr\"{o}dinger’s scenario. 
    We thus require that the dart can only land directly on one of the axes that defines our space. While this requirement appears odd at this stage, we will see later, that this corresponds to a normal game of darts when the number of dimensions grow.
    \item[Rule 2] When the dart lands on a certain axis, the distance from the origin is noted. This distance, given Eq. (\ref{catstate}), represents one measurement of the quantity  $\alpha$ or $\beta$ depending on the axis the dart lands on. Indeed, as per the discussion that follows Eq. (\ref{cat-vector}), the probability that the dart lands on any given axis is proportional to the length $\alpha$ (or $\beta$). 
    \item[Rule 3] Given Eq. (\ref{norm}), the net probability of the dart landing on either axis is 1.
    \item[Rule 4] Whether the dart lands on the horizontal or vertical axes is determined by the magnitude of $\alpha$ and $\beta$. The greater a particular value is, more likely that the dart lands on that specific axis. More specifically, these likelihoods are $\alpha^2$ and $\beta^2$ as noted above.
\end{packed_item}
Without loss of generality, our axes were chosen as the horizontal and vertical axis, and we labeled these as ``dead'' and ``alive''. This is consistent 
with the original thought experiment where, when we observe the cat, it must be either dead or alive. 
Thus, our dart throw must also provide us with one of the two results. 

Furthermore, 
the result of a single throw of dart is completely random and can result in a dead state or an alive state, with probabilities $\alpha^2$ and $\beta^2$. Thus, with a single dart throw, we don’t know much about the state of the cat prior to the throw. 
However, with multiple such dart throws, 
the ratio of the number of darts landing
on the ``alive'' axis to the number of darts landing on the ``dead'' axis would approach the ratio of the probabilities, 
namely, $\alpha^2/\beta^2$. 

\subsection{Game of darts in $N$-dimensions may be interpreted as a Monte Carlo problem}
\label{n-Dim}

We may 
express a function by enumerating its values at every single point on the number-line as
\begin{align}
\begin{pmatrix}
  y_1 \\
  y_2 \\
  \vdots \\ 
  y_n
\end{pmatrix}
\label{n-vector}
\end{align}
where the value, $y_i$, represents the function’s value at $x_i$. This infinitely long vector is not dissimilar to that in Eq. (\ref{cat-vector}).

\begin{figure}[tbp]
\includegraphics[width=0.75\columnwidth]{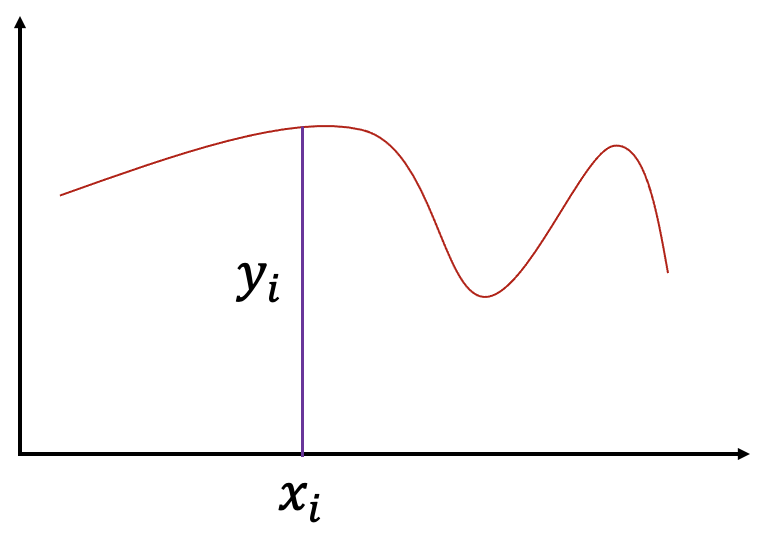}
\caption{\label{fig:MC} An alternative imagery of the dartboard problem for Eq. (\ref{n-vector}). The probability of a dart landing at $x_i$ is proportional to $y_i$. }
\end{figure}
The rules for the dartboard game in Section \ref{darts} when applied to this vector space, leads to an appealing interpretation. 
In our game of darts discussed above, each value along the vector in Eq. (\ref{cat-vector}) corresponds to one specific component that dictates the probability that the dart lands on a given axis. The same set of rules when applied to Eq. (\ref{n-vector}) yields a game of darts with $n$ orthogonal dimensions where the probability of the dart landing on {\em an axis labeled as} $x_i$, is give by $y_i$. 
However, since Eq. (\ref{n-vector}) also represents some function, 
we can re-express this multidimensional game of darts as being played on a two-dimensional coordinate system (see Figure \ref{fig:MC}), where
every throw of darts can land {\em anywhere along the horizontal axis}, with the probability of a given throw landing on a specific point along the horizontal axis being given by the square of the vertical axis value at that point as shown in Figure \ref{fig:MC}! This leads us to a natural interpretation that appears similar to a Monte Carlo problem. 

\section{Functions in Fourier space lead to {\em complementary} Fourier dartboards}
Now, consider two real functions: $\cos(kx)$ and $\sin(mx)$, such that $k$ and  $m$ are integers. Since we’ve established above that we can treat these functions as vectors, as in Eq. (\ref{n-vector}), that is, 
\begin{align}
\cos(kx) \equiv 
\begin{pmatrix}
  \cos(kx_1) \\
  \cos(kx_2) \\
  \vdots \\ 
  \cos(kx_n)
\end{pmatrix}
\label{cos-vector}
\end{align}
and similarly 
\begin{align}
\sin(mx) \equiv 
\begin{pmatrix}
  \sin(mx_1) \\
  \sin(mx_2) \\
  \vdots \\ 
  \sin(mx_n)
\end{pmatrix}
\label{sin-vector}
\end{align}
we can the express the dot product of these vectors as a sum of products, that is, 
\begin{align}
    \int_{-\pi}^{\pi} dx \cos(kx) \sin(mx)
    \label{cosdotsin}
\end{align} 
which is identically zero. Thus, we may say that the sine and cosine functions are essentially vectors that are orthogonal to each other. A similar statement can be made for any two Cosine or any two Sine functions as well, that is
\begin{align}
    \int_{-\pi}^{\pi} dx \cos(kx) \cos(mx) = \delta_{k,m}
    \label{cosdotsin}
\end{align} 
and
\begin{align}
    \int_{-\pi}^{\pi} dx \sin(kx) \sin(mx) = \delta_{k,m}
    \label{cosdotsin}
\end{align} 
where $\delta_{k,m}$ is the Kroneckar delta function.
As a result, we may use a family of Sine and Cosine functions, with different values for $k$ and $m$ in Eqs. (\ref{cos-vector}) and (\ref{sin-vector}) 
 to create a $n$-dimensional vector space where each axis is labelled using Eq. (\ref{cos-vector}) or (\ref{sin-vector}). 
 For example, $\cos(x)$ and $\cos(2x)$ 
define a two-dimensional vector space. 
If we add a third function, such as $\cos(3x)$, we have a three-dimensional
space since $\cos(3x)$ is orthogonal to both $\cos(x)$ and $\cos(2x)$ and so on. 
With $n$ such functions, we have now created an $n$-dimensional space and as we let $n\rightarrow\infty$, we
have an infinite-dimensional vector space.
Thus, in a manner that is complementary to Eq. (\ref{n-vector}) {we may express any function as a linear combination of the vectors that make up our new infinite dimensional basis, which are the
sinusoidal functions in the form} $\cos(kx)$ and $\sin(kx)$, where $k \in \mathbb{Z}$: 
\begin{align}
    f(x) = \sum_{i=0}^\infty \left[ A_i \cos(k_i x) + B_i \sin(k_i x) \right]
    \label{n-wave-function}
\end{align}
or to present this in a fashion complementary to Eq. (\ref{n-vector}):
\begin{align}
\begin{pmatrix}
  A_0 \\  A_1 \\ 
  \vdots \\ 
  A_n \\  B_0 \\ B_1 \\
  \vdots \\ 
 B_n
\end{pmatrix}
\label{n-wave-vector}
\end{align}
which exemplifies the many findings we just made. The coefficients $\left\{ A_n; B_n \right\}$ are evaluated by using the ``dot'' product:
\begin{align}
    A_m = \int_{-\pi}^{\pi} dx f(x) \cos(mx)
    \label{cosdotf}
\end{align} 
As $f(x)$ is made up of a
certain {\em amount} of each wave, and we have shown that these waves are independent of one another (their dot product is 0), taking the dot product of $f(x)$
and one wave component, $\cos(nx)$ or $\sin(nx)$, will yield a numerical value corresponding to the amount of that wave in $f(x)$. 

\begin{figure}[tbp]
\includegraphics[width=0.75\columnwidth]{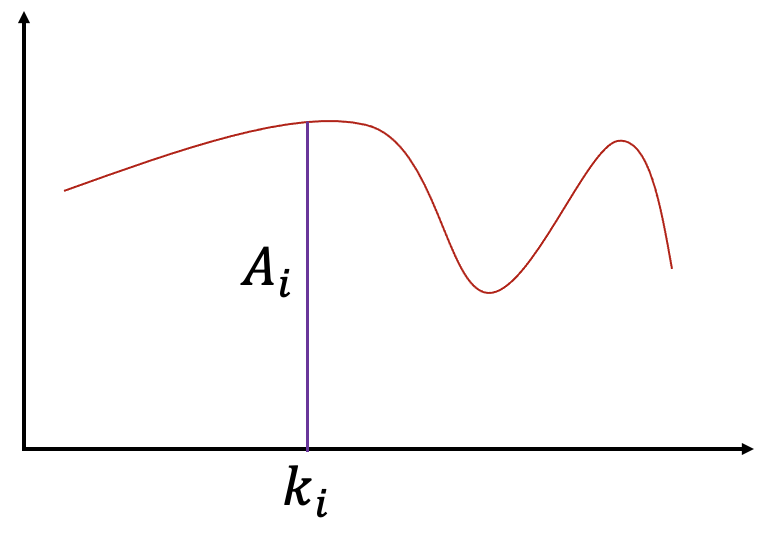}
\caption{\label{fig:kspace} The dartboard problem for Eq. (\ref{n-wave-vector}). The probability of a dart landing at $k_i$ is proportional to $A_i$. A similar figure may be constructed for the Sine functions.}
\end{figure}
\subsection{A complementary $N$-Dimensional Dartboard in Fourier space}
\label{Fourier-darts}
The rules for our game of dart discussed in Section \ref{darts} 
when applied to Eq. (\ref{n-vector}), led to a Monte Carlo interpretation in Section \ref{n-Dim}. 
But we may apply the same rules to Eq. (\ref{n-wave-vector}) as represented by the Figure \ref{fig:kspace}. 
Based on ``Rule 4'' above, this leads to the interpretation that 
the probability of the dart landing on any given axis of our new basis space (or any given point in the horizontal axis of Figure \ref{fig:kspace}) is dictated by the extent to which a specific frequency captured by the sine or cosine wave is present in the function. 
Thus, rather than having just two axes (the “dead” and “alive”
axes from the first example), in Section \ref{n-Dim} and here, we now have an infinite number of orthogonal axes. Each axis is labeled here by a wave
—specifically, its frequency. A point in this space, like in the first example, is simply a linear combination of these axes. From the discussion
of Fourier components, we know that every function can be uniquely represented as a point in this space.

For example, given the function
\begin{align}
    f(x) = a \cos(x) + b \cos(2x) + c \sin(5x)
\end{align}
the probability of our dart throw landing along the $\cos(x)$ axis, or at a point with $k_i=1$ in Figure \ref{fig:kspace}, would be $\frac{a^2}{a^2+b^2+c^2}$. Similarly, the probability of our dart throw landing along the $\cos(x)$ axis, or  at a point with $k_i=2$ in Figure \ref{fig:kspace}, would be $\frac{b^2}{a^2+b^2+c^2}$, and so on. Notice how in the cat scenario, the calculation of the probabilities for landing on a specific axis is the same as the calculation above.
The “alive” probability would be $\frac{\alpha^2}{\alpha^2+\beta^2}$, and the “dead” probability would be $\frac{\beta^2}{\alpha^2+\beta^2}$. However, due to the normalization condition, where $\alpha^2+\beta^2=1$, this simplifies to $\alpha^2$ and $\beta^2$ respectively.

\section{The uncertainty dilemma between these complementary dartboards}
Let us go back to Figure \ref{fig:MC} and imagine that our function $f(x)$, represented by Eq. (\ref{n-vector}), is an exact Cosine wave, say $\cos(x)$. In such a situation, Eq. (\ref{n-wave-vector}) would have exactly one value, $A_1=1$, and all other values in Eq. (\ref{n-wave-vector}) would be identically zero. Consequently, while Figure \ref{fig:MC} would display a wave with a single frequency, where the dart may land anywhere along the horizontal axis, the complementary form of Figure \ref{fig:kspace} would yield a single peak at $k_1=1$, zero everywhere else, thus yielding a probability of "1" that the dart lands at one precise point. {\em Thus, while the outcome of every throw of dart is precisely dictated, with certainty in Figure \ref{fig:kspace}, the corresponding outcome of a single throw of darts may land anywhere along the horizontal axis in Figure \ref{fig:MC}, and hence is uncertain}!

Of course the description above is general, and does not really only apply to quantum mechanics. But if we were equipped with De Broglie's wave-particle duality, where the momentum of a particle is dictated by the frequency associated to its wave-nature, it is clear from the above discussion that the function, $\cos(x)$ has a precisely defined momentum. However, given that the game of darts in Figure \ref{fig:MC} may land anywhere along the horizontal axis, the corresponding position is infinitely uncertain!
Thus one may say, that using the game of darts, we have picturized in Figure \ref{fig:kspace}, a momentum space, and in Figure \ref{fig:MC}, a complementary position space. But our development above did not need the principles of quantum mechanics and are hence expected to be more general. Indeed, it is well-known that a similar time-frequency uncertainty exists in signal processing\cite{Numerical-Recipes}.
\section{Schrodinger’s Cat as a Qubit, darts as measurements} 
Now, let us move on to the connection to quantum bits. Quantum bits, also known as qubits, are similar to the Schr\"odinger’s cat dartboard
example we began with earlier. This time, instead of our two states being ``dead'' and ``alive'', we label the states as $\ket{0}$ and $\ket{1}$, which are again defined to be orthogonal to each other as before. If we define  $\ket{0}$ along the horizontal
axis and $\ket{1}$ along the vertical axis, then, our qubit is similar to our cat-state from Section \ref{cats}. 
We can
express it in an analogous fashion as in Eq. (\ref{catstate}), and
\begin{align}
    \ket{\psi} = \alpha \ket{0} + \beta \ket{1}
    \label{qubitstate}
\end{align}
where $\ket{\psi}$ represents a qubit state with components $\alpha$ and $\beta$ which dictate 
the probabilities of ``collapsing'' the state to the $\ket{0}$ and $\ket{1}$ directions, respectively, exactly as in the game of darts.

Note how qubits differ from standard bits. Standard bits can only be a 0 or a 1. Qubits, on the other hand, can exist in a superposition as in Eq. (\ref{qubitstate}). But, when measured, equivalent to our earlier dart throws, these qubits collapse to a $\ket{0}$ or $\ket{1}$ state.

Now consider two qubits. Since these can both independently reside on their respective $\ket{0}$ and $\ket{1}$ states, the two-qubit system is a four-dimensional space with axes labeled as, 
$\ket{00}$, $\ket{01}$, $\ket{10}$, and $\ket{11}$. How would we
be able to express this pair of qubits? The key is in recognizing that these four states are, similar to (a) the $\ket{0}$ and $\ket{1}$ states for a single qubit, (b) the basis states for the function in Eq. (\ref{n-vector}), and (c) the states in Eqs. (\ref{n-wave-function}) and (\ref{n-wave-vector}), that is, the states $\ket{00}$, $\ket{01}$, $\ket{10}$, and $\ket{11}$ are
orthogonal. Four orthogonal states implies a four-dimensional space, and a state, 
\begin{align}
    \ket{q} = q_{00} \ket{00} + q_{01} \ket{01} + q_{10} \ket{10} + q_{11} \ket{11}
    \label{2qubitstate}
\end{align}
is a point in this four dimensional space. Again all our discussions on the dartboards apply to this state. 
The square of the coefficients, as usual, corresponds to the probability of the pair of qubits collapsing to one specific dimension in this four-dimensional space.
But, upon adding an extra qubit, we have an expanded game of darts. If we were
to add a third then we have eight different possible basis states and in general we have $2^N$ orthogonal dimensions for $N$-qubits. In this fashion the complexity and the ability to store (and process) information grows exponentially from a family of qubits. 

We may also construct an analogy between Eq. (\ref{2qubitstate}) and the state corresponding to two cats, that may be independently ``dead'' or ``alive''. Thus, 
\begin{align}
    \ket{2cats} =& q_{alive_1 alive_2} \ket{alive_1 alive_2} + \nonumber \\ & q_{alive_1 dead_2} \ket{alive_1 dead_2} + \nonumber \\ & q_{dead_1 alive_2} \ket{dead_1 alive_2} + \nonumber \\ & q_{dead_1 dead_2} \ket{dead_1 dead_2}
    \label{2qubit-catstate}
\end{align}
where for example the probability of recovering both cats as living is proportional to $q_{alive_1 alive_2}$. 

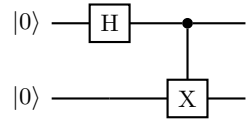
\begin{figure}[t]
\begin{quantikz}[]
   \lstick{$\ket{0}$} & \gate{\text {H}} &  \ctrl{1} & \qw \\
   \lstick{$\ket{0}$} & \qw &  \gate{{\text {X}}} & \qw 
\end{quantikz}
\caption{\label{fig:Bell}A quantum circuit to create the Bell state in Eq. (\ref{2qubitBellstate}). An initial state, $\ket{0}\otimes\ket{0}$ is propagated through a Hadamard transform of qubit one to yield, $\left\{ \frac{1}{\sqrt{2}}\left[\ket{0}+\ket{1}\right]\otimes\ket{0}\right\}$. Following this, application of a CNOT gate yields the state in Eq. (\ref{2qubitBellstate}).}
\end{figure}
\section{Bell states using ``entangled'' darts}
We now introduce a sophisticated concept in quantum information known as entanglement, which is used in the context of quantum teleportation. We develop a dartboard analogue to this concept. For this we first introduce a Bell state\cite{Nielsen-Chuang} which is created from a subset of the four vectors in Eq. (\ref{2qubitstate}), namely
\begin{align}
    \ket{q_B} = \frac{1}{\sqrt{2}} \left[ \ket{00} + \ket{11} \right]
    \label{2qubitBellstate}
\end{align}
Thus by comparison with Eq. (\ref{2qubit-catstate}), when cat-1 is recovered to be alive, this is also the case for cat-2 and vice versa. Thus, in some sense, the outcomes of the throw of darts corresponding to the two cats, has not become ``entangled''. That is the outcome from the first dart throw, also dictates and influences the second dart. 
A quantum circuit that creates such a state is shown in Figure \ref{fig:Bell}. How does one create a picture of this  using our dartboard analogy? 

Let us imagine two darts, that as we will discover, behave in a very peculiar fashion. The two wires in Figure \ref{fig:Bell} represent the two different dartboards where these darts are supposed to land. We will also assumes, as done in Eq. (\ref{2qubit-catstate}), that there exists two cats, whose lives are influenced by the outcome from these two darts and their respective boards. 
Dartboard-1, the top wire in Figure \ref{fig:Bell}, is pre-processed through a transformation given by the Hadamard transform $H$ in Figure \ref{fig:Bell}. This Hadamard transform simply rotates the state that dictates the dart landing probabilities for Dartboard-1 by 45 degrees. So our Dartboard-1 will now present dart-1 with equal probabilities of landing along the horizontal and vertical axis. Thus, in a sense, going back to our cat states in Eq. (\ref{catstate}), the Hadamard transform has created a state that looks like, $\frac{1}{\sqrt{2}} \left[\ket{alive} \pm \ket{dead} \right]$ and notice then by comparison with Eq. (\ref{catstate}), that both $\alpha^2$ and $\beta^2$ are equal. Thus after the Hadamard transform, Dart-1 is free to land on either axis, with no bias. 

Let us now presume that Dart-2 begins, with a dartboard where the dart throws are dictated by the probability, using the cat-state analogy, $\alpha=0$ and $\beta=1$. Thus we know with certainty that the state that dictates Dart-2 is, for example, the $\ket{0}$ state, or the cat represented here is, for example $\ket{alive}$ with probability 1. Thus without influence from Dart-1, Dart-2 will always provide an outcome where the cat is ``alive''.

Now something very strange happens as a result of the operation \\
\begin{center}
\begin{quantikz}[]
   \qw & \ctrl{1} & \qw \\
   \qw &  \gate{{\text {X}}} & \qw 
\end{quantikz}
\end{center}
\noindent in Figure \ref{fig:Bell}. The two darts are not independent anymore. When Dart-1 lands on the state $\ket{0}$, that is finds that the corresponding cat is ``alive'', Dart-2, dictated by the probabilities that control Dartboard-2 discussed in the previous paragraph, recovers an an $\ket{alive}$ or $\ket{0}$ too. But, when Dart-1 lands on the state $\ket{1}$ that is finds that the corresponding cat is ``dead'', Dart-2, promptly {\em flips} the probabilities in the state that dictates Dartboard-2 and also recovers a $\ket{dead}$ or $\ket{1}$ state. Thus the behavior or Dart-2 is not independent, from that of Dart-1, and this property is called entanglement. 
This bizarre result is one of the major hallmarks of quantum mechanics and is thought to be central to a presumed advantage that a computer created from the principles of quantum mechanics may have over those that we currently use.


\section{Conclusion}
We develop an approach to discuss the fundamental underpinnings of quantum mechanics and quantum computing using the concept of dart throws onto the two axis of a two-dimensional space. The concept of qubits and measurement is related to these dart throws and the process, when generalized, yields visual descriptions of uncertainty and Fourier transforms. We follow the prescriptions of this game and develop a way to relate these concepts to fundamental notions in quantum information storage and processing, including analogues to quantum circuits and Bell states. 

\section*{Acknowledgments} This research was supported by the National Science
Foundation grants OMA-1936353 and CHE-2102610 to SSI.

\providecommand{\latin}[1]{#1}
\providecommand*\mcitethebibliography{\thebibliography}
\csname @ifundefined\endcsname{endmcitethebibliography}
  {\let\endmcitethebibliography\endthebibliography}{}


\end{document}